\documentclass[aps,pre,reprint,superscriptaddress,longbibliography]{revtex4-1}
\usepackage{scrextend} %,wasysym}
\usepackage{amsmath,amssymb,graphicx,subfigure,color,times,tabularx,hyperref,fancyhdr}
\usepackage{tcolorbox,xcolor}
\usepackage{esint}
\begin{document}
\title{The Finite Coulomb Lattice Sum: A Resolution of Conditional Convergence through Exact Shape and Size}
\author{Yang He}
\affiliation{Key Laboratory of Laser \& Infrared System of Ministry of Education, Shandong University, Qingdao 266237, P. R. China}
\affiliation{Qingdao Institute for Theoretical and Computational Sciences (QiTCS), Center for Optics Research and Engineering, Shandong University, Qingdao 266237, P. R. China}
\author{Zhonghan Hu} \email{zhonghanhu@sdu.edu.cn}
\affiliation{Key Laboratory of Laser \& Infrared System of Ministry of Education, Shandong University, Qingdao 266237, P. R. China}
\affiliation{Qingdao Institute for Theoretical and Computational Sciences (QiTCS), Center for Optics Research and Engineering, Shandong University, Qingdao 266237, P. R. China}
%\date{\today}

\begin{abstract}
This work examines conditionally convergent Coulomb lattice sums under periodic boundary conditions.
The recently developed finite lattice sum cleanly decomposes the series into three distinct components: a periodic bulk term $\nu_{\rm pbc}$, a shape-dependent non-periodic boundary term $\nu_{\rm b}$, and a finite-size correction term $\nu_{\rm corr}$.
This rigorous formulation explicitly parameterizes the geometry of a finite lattice by its exact shape and size and takes an effective pairwise form.
We analyze it in detail and compare it with various derivations of lattice sums in the literature.
Perspectives on future applications are discussed, including analytical developments for arbitrarily shaped crystals and numerical mesh-type algorithms for condensed matter simulations.
\end{abstract} \maketitle
%%%%%%%%%%%%%%%%%%%%%%%%%%%%%%%%%%%%%%%%%%%%%%%%%%%%%%%%%%%%%%%%%%%%%%%%%%%%%%%%%%%%%%%%%%%%%%%%%%%%%%%%%%%%%%%%%%%%%%%%%%%%%%%%%%%%%%%%%%%%%%%%%%%%%%%%%%%%%%%%  Introduction
Simulations of condensed matter systems often employ periodic boundary conditions (PBCs), in which a unit cell containing charges (nuclei, ions, electrons, etc.) is surrounded by an infinite array of periodic replicas. This approach is widely used in molecular dynamics simulations of liquids and biomolecular systems\cite{Allen_Tildesley2017}, electronic band structure calculations of solid-state materials\cite{Singleton2001}, and quantum Monte Carlo simulations of crystals\cite{Cazorla_Boronat2017}.

Under PBCs, the key calculation involves determining the electrostatic potential at a reference charge arising from an infinite lattice of identical unit cells.
The computation of this Coulomb lattice sum, historically referred to as the Madelung series, is a century-old problem\cite{Madelung1918,Ewald1921,Born_Huang1954}. The resulting potential, scaled by the nearest-neighbor interaction, defines the dimensionless Madelung constant essential for characterizing the influence of the surrounding lattice.

Summing Coulomb interactions across an infinite lattice presents a fundamental mathematical challenge: the series is only conditionally convergent, meaning its value depends explicitly on the order of summation.
This subtlety was likely not fully appreciated at the time Ewald developed his celebrated technique (see $\nu_{\rm pbc}$ in the Appendix)\cite{Ewald1921}.
The importance of a rigorous mathematical treatment of such series was first recognized by Redlack and Grindlay\cite{Redlack_Grindlay1972,Redlack_Grindlay1975}, who pointed out that the lattice sum must include an additional term --- beyond the standard periodic expression of Ewald --- that depends on the macroscopic shape of the crystal. 
Ever since, significant physical insight has been gained through more rigorous, generalized, and alternative derivations of this shape-dependent term 
(see, e.g., Refs.
\cite{Tupizin_Abarenkov1977,Felderhof1980,DeLeeuw_Smith1980,DeLeeuw_Smith1980a,Smith1981,Smith_Perram1983,Hansen1986,Smith1988,Pisani1988,Saunders1992,Roberts_Schnitker1994,Makov_Payne1995,Fraser_Williamson1996,Hunenberger1999,Kantorovich_Tupitsyn1999,Heinz_Hunenberger2005,Herce_Darden2007,Smith2008,Ballenegger2014,Hu2014ib,Pan_Hu2017,Zhao_Hu2025}; this list is by no means exhaustive).

Recently, we introduced a rigorous formulation for finite lattice sums by explicitly parameterizing the exact shape and size of a crystal\cite{Zhao_Hu2026}.
This formulation applies across all scales, from macroscopic crystals down to just a few replicas of unit cells.
For the first time, it decomposes the term in addition to the Ewald expression into two distinct components: a size-independent boundary term ($\nu_{\rm b}$) and a finite-size correction term ($\nu_{\rm corr}$) that decays monotonically with increasing crystal dimension.
By explicitly incorporating the boundary term alongside the leading-order finite-size correction, we obtain a rapidly convergent direct-summation scheme.
This method achieves exceptional accuracy (errors $\sim 10^{-4}$) even for the smallest supercells of $3^3=27$ unit cells, enabling hands-on calculations of Madelung constants for a wide range of ionic crystals, including CsCl, NaCl, ZnS, CaF$_2$, and CaTiO$_3$\cite{Zhao_Hu2026}. 

In this Perspective, we analyze the rigorous separation of the total finite lattice sum into these three components --- $\nu_{\rm pbc}$, $\nu_{\rm b}$, and $\nu_{\rm corr}$ --- in an intuitive manner without involving complex mathematical derivations. 
The analysis elucidates the physical significance of these components in light of historical theoretical developments and suggests promising future research directions in connection with recent advances in the literature.
Complete formulations of lattice sums with an arbitrarily shaped unit cell are given in the appendix for further reference.
%%%%%%%%%%%%%%%%%%%%%%%%%%%%%%%%%%%%%%%%%%%%%%%%%%%%%%%%%%%%%%%%%%%%%%%%%%%%%%%%%%%%%%%%%%%%%%%%%%%%%%%%%%%%%%%%%%%%%%%%%%%%%%%%%%%%%%%%%%%%%%%%%%%%%%%%%%%%%%%%  ------------------- Shape and Size -------------
\begin{figure*}[!htb]\centerline{\includegraphics[width=16cm]{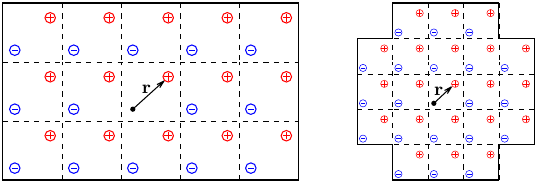} }           %%%%%%%%%%%%%%%%%%%%%%%%%%%%------------ Figure 1 ------------------------------------
\caption{Schematic representations of two finite 2D centro-symmetric crystals. Solid lines delineate the exact crystal boundaries, and black dots indicate negative ions located in the central unit cells.
Each cell contains two unit charges ($\pm q = \pm 1$) separated by a displacement vector ${\mathbf r}$.
The crystal on the left has dimensions of $5\times 3$ unit cells, while the one on the right measures exactly $3\times 3 + 3\times 1$, or equivalently $5\times 5 - 1\times 1$ unit cells --- in contrast to earlier approximate descriptions that treated the shape as circular (e.g., Ref.\cite{DeLeeuw_Smith1980,Smith1981}).
The corresponding next-larger crystals in each sequence would be $15\times 9$ for the left, and $9\times 9 + 9\times 3$ (or $15\times 15 - 3\times 3$) for the right.}\label{fig:crystals}\end{figure*}

\begin{tcolorbox}[colback=yellow!30, colframe=black!75,colframe=white,boxrule=0pt]
{\Large Exact geometric scaling on a discrete lattice requires regular crystals to be parameterized by a triplet of integers and composite or irregular shapes by an extended set of integers.}
\end{tcolorbox}

%\textcolor{blue}{
When discussing conditional convergence, it is tempting to draw an analogy with the well-known one-dimensional (1D) alternating harmonic series found in mathematical textbooks:
\begin{equation}  S_{\rm 1D} = 1 - \frac{1}{2} + \frac{1}{3} - \frac{1}{4} + \cdots,  \end{equation}
which yields different results depending on how the terms are grouped. Naturally, 
\begin{equation} \lim_{p\to\infty} \sum_{n=1}^p \left[\frac{1}{2n-1} - \frac{1}{2n} \right] = \ln 2. \end{equation}
Alternatively, one might break the natural order and rearrange the terms to obtain
\begin{equation} \lim_{p\to\infty} \sum_{n=1}^p \left[\frac{1}{2n-1} - \frac{1}{4n-2} - \frac{1}{4n} \right] = \frac{1}{2} \ln 2 . \end{equation}
However, this specific type of conditional convergence involving a broken order of terms is not the focus here.
In fact, $S_{\rm 1D}=\ln 2$ corresponds to the Coulomb lattice sum for a pair of unit charges separated by unity in a 1D periodic system with a unit cell of length $2$.
As such, the 1D Coulomb lattice sum does not give rise to conditional boundary terms in a physical sense\cite{Hu2014ib}.
%}

%\textcolor{blue}{
A more appropriate analogy for the shape-dependent boundary term is provided by the prototypical alternating series\cite{Zhao_Hu2025}:
\begin{equation} S = 1 - 1 + 1 - 1 + \cdots. \label{eq:s0} \end{equation}
Two summation conventions without breaking the natural order arise:
\begin{equation} S_{-} = (1-1) + (1-1) + \cdots = 0, \end{equation}
which corresponds to a sequence ending in $-1$; and 
\begin{equation} S_{+} = 1 + (-1 + 1) + (-1 + 1) + \cdots = 1, \label{eq:splus} \end{equation} which corresponds to a sequence ending in $+1$.
These two summations can be unified as $S_{\mp} = S_{\rm bulk} \mp 1/2$.
This formulation separates the series into two distinct contributions: a bulk component that satisfies intrinsic periodicity ($S = 1 - S$, yielding $S_{\rm bulk} = 1/2$), and a boundary term ($\mp 1/2$) that reflects the choice of the final term in the series.
%}

The mathematical complexity of a Madelung series arises because the lattice summation extends over an infinite number of lattice vectors in three dimensions (3D), rather than in one dimension.
Unlike well-defined 1D series, a 3D lattice sum depends critically on the order in which terms are accumulated.
Physically, different summation orders correspond to different asymptotic boundary as the crystal grows to infinity, yielding distinct limiting values. 

A rigorous treatment must therefore eliminate this ambiguity by recasting the triple sum into a one-dimensional sequence of partial sums, where each term represents a finite, unconditionally convergent lattice sum.
Constructing such a sequence is, however, nontrivial. 
For an arbitrary lattice, we first consider a family of finite, centro-symmetric crystals parameterized by a fixed triplet of non-negative integers ${\mathbf s}=(s_1, s_2, s_3)$ and an index $p \geqslant 0$.
The crystal dimensions (in unit cells) are $(2p+1)(2s_1+1)\times (2p+1)(2s_2+1)\times (2p+1)(2s_3+1)$, with the triplet $(2s_1+1,2s_2+1,2s_3+1)$ constrained to be coprime to ensure an irreducible aspect ratio.
Advancing through the sequence by increasing $p$ uniformly expands all linear dimensions, thereby preserving both the centro-symmetric geometry and the exact aspect ratio.
In this setup, the scalar variable $p$ controls the size, while the vector ${\mathbf s}$ fixes the crystal shape; for an orthorhombic lattice, this regular shape corresponds to centro-symmetric rectangular prisms\cite{Zhao_Hu2025,Zhao_Hu2026}.
As a two-dimensional illustration, the left panel of Fig.~\ref{fig:crystals} depicts a cross-section for $p=0$, $s_1=2$, and $s_2=1$, with each unit cell containing a pair of unit charges ($\pm q = \pm 1$) separated by a displacement vector ${\mathbf r}$.

Historically, spherical or ellipsoidal boundaries have been widely employed to approximate large finite crystals\cite{DeLeeuw_Smith1980,Smith1981,Kantorovich_Tupitsyn1999,Smith2008}.
However, such continuous curved geometries cannot be exactly realized on a discrete lattice while maintaining a strict homothetic expansion in a sequence; the spherical or ellipsoidal geometry remains inherently approximate due to the staircase truncation of unit cells at the boundary.
To maintain exact geometric scaling on a discrete lattice, irregular or composite shapes must instead be constructed from combinations of regular shapes.
In such cases, the shape parameter ${\mathbf s}$ generalizes to an extended set of integers that uniquely define the composite geometry.
The right panel of Fig.~\ref{fig:crystals} illustrates a cross-section of one such composite crystal, whose exact boundary is rigorously defined by the union of two distinct regular shapes.

\begin{tcolorbox}[colback=yellow!30, colframe=black!75,colframe=white,boxrule=0pt]
{\Large The definition of exact shape and size for a finite crystal enables a clear separation of boundary and finite-size effects.}
\end{tcolorbox}
Let ${\mathcal L}(p|{\mathbf s})$ denote the set of all lattice vectors defining the finite crystal of size $p$ and shape ${\mathbf s}$.
The index ${\mathbf n}\in {\mathcal L}(p|{\mathbf s})$ enumerates these vectors, with ${\mathbf n}={\mathbf 0}$ corresponding to the central unit cell.
The electrostatic potential experienced by the target negative ion ($-q=-1$) in the central cell (see Fig.~\ref{fig:crystals}) is then given by
\begin{equation} \nu({\mathbf r},p|{\mathbf s}) = \frac{1}{r} + \sum_{ {\mathbf n}\neq {\mathbf 0} }^{{\cal L}(p|{\mathbf s})} \left[ \frac{1}{\left| {\mathbf r} + {\mathbf n} \right | }  - \frac{1}{n} \right] , \label{eq:dir}   \end{equation}
which accounts for the Coulomb interactions with all other charges in the finite crystal.
By symmetry, $\nu({\mathbf r},p|{\mathbf s})$ is equal in magnitude but opposite in sign to the electrostatic potential experienced by the central positive ion.
Consequently, it exactly represents the negative of the energy of the central dipole (dipole moment $q{\mathbf r}$) formed by the charge pair ($\pm q = \pm 1$).

Because the order of summation does not matter in finite summations, summing over all lattice vectors in each finite crystal of the defined sequence uniquely specifies a mathematical sequence of finite lattice sums: $\nu({\mathbf r},0|{\mathbf s}), \nu({\mathbf r},1|{\mathbf s}), \dots$.
For any given unit cell, this mathematical sequence converges unambiguously to $\nu({\mathbf r},\infty|{\mathbf s})$, which corresponds to the chosen infinite lattice sum characterized by the specific shape prescribed by ${\mathbf s}$.
For a finite crystal of size $p$ in the sequence, $\nu({\mathbf r},p|{\mathbf s})$ deviates from $\nu({\mathbf r},\infty|{\mathbf s})$ by a finite-size correction (corr) term
\begin{equation} \nu_{\rm corr}({\mathbf r},p|{\mathbf s}) =  - \sum_{ {\mathbf n}\notin {\cal L}(p|{\mathbf s}) }^{{\cal L}(\infty|{\mathbf s})} \left[ \frac{1}{\left| {\mathbf r} + {\mathbf n} \right | }  - \frac{1}{n} \right],  \label{eq:corrdef} \end{equation}
where the sum extends over all lattice vectors excluded from the finite crystal but present in the infinite lattice of the same shape.

To derive the leading order term of $\nu_{\rm corr}({\mathbf r},p|{\mathbf s})$, the summand for distant cells ($\left| {\mathbf n} \right| \gg \left| {\mathbf r}\right|$) can be expanded as a multipolar series\cite{Gelle2008,Greengard1988,Zhao_Hu2026}.
Under the exact scaling condition that all linear dimensions along the boundary expand in strict proportion to maintain a fixed geometric ratio independent of $p$, the leading order correction derived from the multipolar series scales as $(2p+1)^{-2}$.
In particular, for a cubic crystal in a cubic lattice of lattice constant $l$ (i.e., ${\mathbf s}=(0, 0, 0)$), the leading order correction for ${\mathbf r}=(x, y, z)$ admits a closed-form expression that scales as ${\mathcal O}(l^{-5})$\cite{Zhao_Hu2026}:
\begin{equation} \nu_{\rm corr}({\mathbf r},p|{\mathbf s}) = \frac{24r^4 - 40(x^4+y^4+z^4)}{9\sqrt{3} (2p+1)^2 l^5} + {\mathcal O}(p^{-4}) \label{eq:corr}. \end{equation}
Notably, the spherical average of this quartic term over the polar and azimuthal angles of ${\mathbf r}$ vanishes\cite{Zhao_Hu2025}.

The imposition of the exact scaling condition is mathematically crucial: it guarantees a well-defined decomposition of the correction term in Eq.~\eqref{eq:corrdef} and significantly simplifies the derivation of its explicit asymptotic form. 
An important open question remains whether specific crystal geometries can be engineered to minimize this finite-size correction for a given unit cell. 
Investigating optimal shapes to accelerate convergence in direct-summation schemes represents a promising direction for future research.

For the infinitely large crystal in the sequence, no finite-size correction is present.
In this limit, $\nu({\mathbf r},\infty|{\mathbf s})$ further decomposes into a periodic bulk term $\nu_{\rm pbc}({\mathbf r})$ and a non-periodic boundary term $\nu_{\rm b}({\mathbf r}|{\mathbf s})$.
The bulk contribution $\nu_{\rm pbc}({\mathbf r})$, originally derived by Ewald\cite{Ewald1921} and commonly referred to as the intrinsic potential, is strictly independent of the shape and size of the crystal and the mode of replication of the unit cell.
In contrast, the sum $\nu_{\rm b}({\mathbf r}|{\mathbf s}) + \nu_{\rm corr}({\mathbf r},p|{\mathbf s})$ --- historically treated collectively as the external potential\cite{Roberts_Schnitker1994} --- varies explicitly with these factors.
Crucially, the rigorous definition of exact shape and size uniquely partitions this external potential into a strictly size-independent boundary term and a vanishing finite-size correction, thereby cleanly isolating the boundary effect arising purely from the nature of long-range Coulomb interaction.

Following the seminal contributions of Redlack and Grindlay\cite{Redlack_Grindlay1972,Redlack_Grindlay1975} and De Leeuw, Perram, and Smith\cite{DeLeeuw_Smith1980,DeLeeuw_Smith1980a,Smith1981,Smith_Perram1983}, the shape-dependent boundary term has been extensively revisited through alternative mathematical treatments\cite{Tupizin_Abarenkov1977,Felderhof1980,Hansen1986,Smith1988,Pisani1988,Saunders1992,Roberts_Schnitker1994,Makov_Payne1995,Fraser_Williamson1996,Hunenberger1999,Kantorovich_Tupitsyn1999,Heinz_Hunenberger2005,Herce_Darden2007,Smith2008,Ballenegger2014,Hu2014ib,Pan_Hu2017,Zhao_Hu2025,Zhao_Hu2026}.
Clearly, if the interactions were short-ranged --- decaying sufficiently fast or strictly vanishing beyond a cutoff --- the lattice sum would converge absolutely, and no shape-dependent boundary contribution would arise.
In such cases, the physics would be fully captured by a periodic bulk term describing local interactions and a finite-size correction accounting for the finite extent of the crystal.

\begin{figure}[!htb]\centerline{\includegraphics[width=7cm]{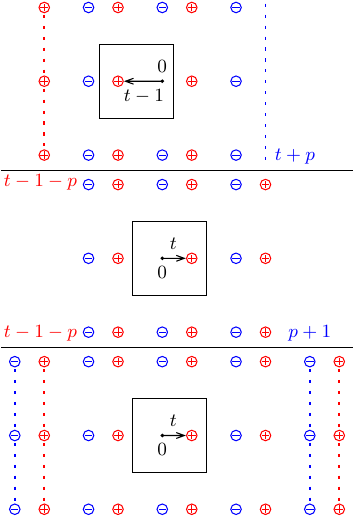} }          %---------------------  Figure -------------------------------------
\caption{
Cross-sectional views of two cubic crystals (top and middle panels) with dimensions $(2p+1)\times(2p+1)\times(2p+1)$ and one cuboid crystal (bottom panel) with dimensions $3(2p+1)\times(2p+1)\times(2p+1)$.
The central unit cells are highlighted by square boxes, and the target ions (dots) are located at the origin. The displacement vectors are $(t-1, 0, 0)$ for the top crystal and $(t, 0, 0)$ for the middle and bottom crystals.
In the limit of large $p$, the electrostatic differences between the two cubic crystals, as well as between the cuboid and cubic crystals, can be approximated either by uniformly charged plates (indicated by dashed lines) or by the differences in surface charge densities of two uniformly polarized continua.
\label{fig:cub3}}\end{figure}

\begin{tcolorbox}[colback=yellow!30, colframe=black!75,colframe=white,boxrule=0pt]
{\Large The physical origin of the boundary term can be intuitively understood by comparing crystals with different displacements or different geometries.}
\end{tcolorbox}
On the basis of the defined exact shape and size, the physical origin of $\nu_{\rm b}({\mathbf r}|{\mathbf s})$ can be grasped intuitively without recourse to complex mathematical formalism.
To illustrate this, consider a cubic unit cell of volume $V=l^3$ and two finite cubic crystals with identical dimensions: $(2p+1)\times(2p+1)\times(2p+1)$ (i.e., ${\mathbf s}_1 =  (0,0,0)$), but differing in their internal displacements: ${\mathbf r}_1 = (t, 0, 0)$ and its periodic translation ${\mathbf r}_2 = (t-1, 0, 0)$. 
These displacements represent two distinct modes of replication, as illustrated in the middle and top panels of Fig.~\ref{fig:cub3}.
With the target ion fixed at the origin, the outermost charged layer of the first crystal consists of a positive surface at $x=t+p$ and a negative surface at $x=-p$.
For the second crystal, the corresponding surfaces are at $x=t-1-p$ (positive) and $x=p$ (negative).
The difference in electrostatic potentials experienced by the target ion, $\nu({\mathbf r}_2,p|{\mathbf s}_1) - \nu({\mathbf r}_1,p|{\mathbf s}_1) $, stems solely from the asymmetric surfaces at $x=t-1-p$ and $x=t+p$.
Furthermore, because the negative surfaces at $x=\pm p$ produce identical potential at the origin, the difference can be equivalently viewed as arising from the difference between the two outermost layers.
As $p\to\infty$, finite-size effects vanish and the bulk terms cancel, yielding $\nu({\mathbf r}_2,\infty|{\mathbf s}_1) - \nu({\mathbf r}_1,\infty|{\mathbf s}_1)  = \nu_{\rm b}({\mathbf r}_2|{\mathbf s}_1) - \nu_{\rm b}({\mathbf r}_1|{\mathbf s}_1)$. 
Additionally, these surfaces approximate uniformly charged square plates (side length $(2p+1)l$ and charge density $\pm 1/l^2$), whose potential at the origin must equal $\nu_{\rm b}({\mathbf r}_2|{\mathbf s}_1) - \nu_{\rm b}({\mathbf r}_1|{\mathbf s}_1)$.
A mathematical treatment of the uniformly charged square plates in the limit $p\to\infty$ has been given\cite{Zhao_Hu2026}.

We further consider a third crystal with dimensions $3(2p+1)\times(2p+1)\times(2p+1)$ (i.e., ${\mathbf s}_2 =  (1,0,0)$) and the same displacement ${\mathbf r}_1$, shown in the bottom panel of Fig.~\ref{fig:cub3}.
The outermost charged layer of this crystal consists of a positive surface at $x=-(3p+1)$ and a negative surface at $x=t+(3p+1)$.
The difference, $\nu({\mathbf r}_1,p|{\mathbf s}_2) - \nu({\mathbf r}_1,p|{\mathbf s}_1) $, arises from the additional asymmetric charged surfaces introduced by the elongated geometry, located at $x=\pm (p+1), t\pm (p+1), \cdots, \pm(3p+1), t\pm (3p+1)$. 
In the continuum limit ($p\to\infty$), the discrete stepped surfaces transition continuously in the $x$ direction. 
The contributions from intermediate asymmetric surfaces effectively average out, leaving only the net difference in surface charge density at the outermost boundaries.
Consequently, $\nu({\mathbf r}_1,\infty|{\mathbf s}_2) - \nu({\mathbf r}_1,\infty|{\mathbf s}_1)=\nu_{\rm b}({\mathbf r}_1|{\mathbf s}_2) - \nu_{\rm b}({\mathbf r}_1|{\mathbf s}_1)$ is again accounted for entirely by the difference in the outermost layers between the two crystals.

The above geometric analysis demonstrates that the boundary term $\nu_{\rm b}$ is governed exclusively by the surface charges at the outermost layer.
Since these surface charges arise from a uniform distribution of dipole moments (one per volume $V$), $\nu_{\rm b}({\mathbf r}|{\mathbf s})$ is related to the interaction energy of the central dipole with the uniform polarization.
This continuum picture naturally yields the general integral representation\cite{Smith2008,Ballenegger2014,Zhao_Hu2025}:
\begin{equation} \nu _{\rm b}({\mathbf r}|{\mathbf s}) = \frac{1}{2V} \int_{\Omega(p|{\mathbf s})} d{\mathbf x} \, \left({\mathbf r}\cdot \nabla_{\mathbf x}\right)^2\frac{1}{\left| {\mathbf x}\right|}, \label{eq:b} \end{equation}
where $\Omega(p|{\mathbf s})$ denotes the volume of the finite crystal.
As expected, although defined over a finite domain, the integral remains independent of $p$ under exact geometric scaling.
Thus, comparing crystals with different displacements or different geometries directly reveals how long-range interactions generate non-periodic shape-dependent boundary effects, providing clear physical insight without complex formalism.

For a rectangular prism with side lengths $a$, $b$, and $c$, its volume and space diagonal are $\Omega(p|{\mathbf s}) = abc$ and $d=\sqrt{a^2+b^2+c^2}$, respectively.
Explicit evaluation of the integral in Eq.~\eqref{eq:b} yields a closed-form quadratic expression for the boundary term in terms of arctangent functions\cite{Zhao_Hu2026,Pan2017}:
\begin{equation} \nu_{\rm b}({\mathbf r}|{\mathbf s}) = -\frac{4}{V} \left(x^2{\rm atan}\frac{bc}{ad} + y^2{\rm atan}\frac{ac}{bd} + z^2{\rm atan}\frac{ab}{cd} \right). \label{eq:br} \end{equation}
This expression is, so far, the most general closed-form result available for the boundary term.
%Deriving analogous exact expressions for crystals with lower symmetry remains an important objective for future research.
As a natural extension of this work, deriving closed-form analytical expressions for the boundary term in arbitrary unit cell geometries, particularly for low-symmetry triclinic crystal systems, remains an important direction for future research.

For the special case of a cubic crystal ($a=b=c$), this expression simplifies dramatically
\begin{equation} \nu_{\rm b}({\mathbf r}|{\mathbf s}) = - 2\pi r^2/(3V). \label{eq:bc} \end{equation}
Eqs.~\eqref{eq:br} and~\eqref{eq:bc} enable explicit verification that the boundary term derived from the uniformly charged plates is equivalent to that obtained from the dipole-dipole interaction energy.
This consistency was confirmed for the crystal geometries shown in Fig.~\ref{fig:cub3} \cite{Zhao_Hu2026}.

With the present exact shape and size defined, more rigorous, generalized, and alternative derivations of the finite lattice sum are expected in the future.
There are already various expressions of the boundary term alternative to Eq.~\eqref{eq:b}.
For example, in reciprocal space, it can be expressed as the ${\mathbf k}=(k_x, k_y, k_z) \to 0$ limit in the reciprocal space contribution of the usual Ewald formulation (see $\nu_{\rm F}$ of Eq.~\eqref{eq:e3dtf2} in the Appendix)
\cite{Hu2014ib,Yi_Hu2015,Yi_Hu2017pairwise,Zhao_Hu2025}:
\begin{equation} \nu_{\rm b}({\mathbf r}|{\mathbf s}) = -\frac{2\pi}{V} \lim_{ {\mathbf k}| {\mathbf s} \to {\mathbf 0} } \frac{\left( {\mathbf k}\cdot {\mathbf r} \right)^2}{\left| {\mathbf k} \right|^2}, \end{equation}
where the limit is taken such that the wavevector ${\mathbf k}$ approaches zero along a direction encoded in ${\mathbf s}$.
The limit corresponds to an asymptotic shape of a macroscopic crystal; hence, this term is also called the infinite boundary term\cite{Hu2014ib}.
This representation makes it evident that $\nu_{\rm b}({\mathbf r}|{\mathbf s})$ is always nonpositive and that its spherical average over the polar and azimuthal angles of ${\mathbf r}$ coincides with Eq.~\eqref{eq:bc}, independent of the shape.
By symmetry, $\nu_{\rm b}({\mathbf r}|{\mathbf s})$ is even in $x$, $y$ and $z$ for the defined highly symmetric shape (cube and sphere); consequently, cross terms such as $xy$, $yz$, etc., do not appear.
A simple formal analysis by choosing the orientation of ${\mathbf k}$ can often yield much information. 
For instance, Eq.~\eqref{eq:bc} can be obtained either by choosing $k_x=k_y=k_z$ and imposing the vanishing of cross terms, or by carrying out a spherical average over the polar and azimuthal angles of ${\mathbf k}$\cite{Hu2014ib}.
Furthermore, it is easy to make connections with 2D Ewald methods\cite{Parry1975,Heyes_Clarke1977,DeLeeuw_Perram1979}, where the boundary term (the limit of the corresponding reciprocal space vector to zero) exists but does not depend on the shape (orientation) any more\cite{Pan_Hu2014,Hu2014ib,Pan_Hu2015,Pan2017,Gan2025,Gao_Gan2025}.

%%%%%%%%%%%%%%%%%%%%%%%%%%%%%%%%%%%%%%%%%%%%%%%%%%%%%%%%%%%%%%%%%%%%%%%%%%%%%%%%%%%%%%%%%%%%%%%%%%%%%%%%%%%%%%%%%%%%%%%%%%%%%%%%%%%%%%%%%%%%%%%%%%%%%%%%%%%%%%%%  ------------------- boundary term 
\begin{tcolorbox}[colback=yellow!30, colframe=black!75,colframe=white,boxrule=0pt]
{\Large Explicit treatment of boundary and finite-size corrections enables direct summation over minimal supercells to yield highly accurate bulk energies.}
\end{tcolorbox}

In a many-body charged system, the two-particle potential $\nu$ and its three components --- $\nu_{\rm pbc}$, $\nu_{\rm b}$ and $\nu_{\rm corr}$ --- serve as effective pairwise interactions.
Consider a unit cell (simulation box) consisting of $N$ ions with charges $q_j$ located at positions ${\mathbf r}_j$ for $j=1,2,\dots,N$.
The electric potential experienced by the $i$-th ion in the central unit cell is then given by\cite{Zhao_Hu2025,Zhao_Hu2026}
\begin{equation} \phi_p(i|{\mathbf s}) = \sum_{j=1,j\neq i}^N  q_j \nu({\mathbf r}_j-{\mathbf r}_i,p|{\mathbf s}) \label{eq:phi}. \end{equation}
The corresponding electrostatic energy of the central unit cell is\cite{Hu2014ib,Yi_Hu2017pairwise,Zhao_Hu2025,Zhao_Hu2026}:
\begin{equation} {\mathcal U}(p|{\mathbf s}) = \frac{1}{2}\sum_{j=1}^N q_j \phi_p(j|{\mathbf s}) = \sum_{i<j} q_i q_j \nu({\mathbf r}_j-{\mathbf r}_i,p|{\mathbf s}).  \end{equation}

In the literature\cite{DeLeeuw_Smith1980,Smith1981,Ballenegger2014}, the shape-dependent boundary contribution to the energy was commonly found to be expressed in terms of the total dipole moment rather than the pairwise formulation.
Smith's original formulation of the boundary effect is written as\cite{Smith1981}
\begin{equation} J({\mathbf M}) = - \frac{1}{V} \left( C_1 M_x^2 + C_2 M_y^2 + C_3 M_z^2 \right) ,\end{equation}
where $C_\alpha$ with $\alpha=1,2,3$ are coefficients related to the macroscopic shape, and the total dipole moment is defined as
\begin{equation} {\mathbf M} = \left(M_x,M_y,M_z\right) = \sum_{j=1}^N q_j {\mathbf r}_j .  \end{equation}
$J({\mathbf M})$ is related to the pairwise formulation through
\begin{equation} \sum_{i<j}^N q_iq_j \nu_{\rm b}({\mathbf r}_i - {\mathbf r}_j|{\mathbf s}) = J({\mathbf M}) - \frac{q_{\rm tot}}{V}\sum_{j=1}^N q_j |{\mathbf r}_j|^2 ,\end{equation}
where the boundary term is of the type in Eq.~\eqref{eq:br}:
\begin{equation} \nu_{\rm b}({\mathbf r}|{\mathbf s}) = \frac{1}{V}\left(C_1x^2 +C_2y^2 +C_3z^2\right), \end{equation}
and $q_{\rm tot} = \sum_{j=1}^N q_j$ is the total charge. 
For a neutral system ($q_{\rm tot}=0$), the second term vanishes, and the pairwise sum reduces exactly to $J({\mathbf M})$.

%\textcolor{blue}{
Of these three contributions to the electrostatic energy, $\nu_{\rm pbc}$ constitutes the usual Ewald sum with the tinfoil boundary condition (see Eqs.~\eqref{eq:e3dtf0} to~\eqref{eq:ew}). 
It captures the intrinsic electrostatic correlations and is widely used in simulations of homogeneous systems with vanishing permanent charge distribution, as well as in electronic structure calculations. 
In contrast, the boundary term $\nu_{\rm b}$ governs surface effects, directly influencing charge fluctuations\cite{Felderhof1980} and the response to external fields that induce inhomogeneities\cite{Caillol1994,Pan_Hu2017,Hu2022}.
Recently, it was found that the tinfoil boundary condition stabilizes polar structures of ice and may not be appropriate for the ice VI--XV transition\cite{Delben_Slater2014}.
This subtle issue deserves further study, although the general consequences of different boundary terms on charge correlations, fluctuations of dipole moments, and the calculation of dielectric constants have been known\cite{DeLeeuw_Smith1980,Caillol1992,Caillol1994}.
%}

%\textcolor{blue}{
In the special case of planar geometry, where $a$ and $b$ in Eq.~\eqref{eq:br} are fixed and both $c/a = 0$ and $c/b = 0$, we have $C_1 = C_2 = 0$ and $C_3 = -2\pi$.
This dipole term was first discussed by Smith \cite{Smith1981} and later used together with the traditional Ewald sum to treat electrostatic in a slab geometry with 2D periodicity\cite{Yeh_Berkowitz1999,Yeh_Wallqvist2011}.
In this approach, a regular Ewald3D method is performed for a 3D periodic system with a large empty space in the $z$ direction, and then corrected with the dipole term accounting for the boundary effect. 
From a mathematical viewpoint of the 2D Ewald method for slab geometry, approximating the Ewald2D formula using the Ewald3D formula---by extending artificial periodicity to the direction perpendicular to the slab via a large empty space between adjacent layers of images---requires an extra layer correction term to account for the presence of these artificial layers\cite{Arnold_Holm2002b,Pan_Hu2014,Gao_Gan2025}.
However, this layer correction term is usually very small and can often be neglected when the empty space is sufficiently large\cite{Yeh_Berkowitz1999,Arnold_Holm2002b,Pan_Hu2014}.
More recently, it has been shown that the dipole correction must be used with caution when simulations are performed under constant electric potential conditions, rather than the standard constant-charge condition\cite{Ahrens-Iwers2021}.
%}

Despite this elegant connection to the total dipole moment, the pairwise sum is more fundamental because the boundary effect can contribute to the electrostatic potential even when its contribution to the energy vanishes.
For example, consider the CsCl lattice using an Evjen-type unit cell\cite{Evjen1932}: a central cation ($+1$) is surrounded by eight anions, each located at a corner but assigned a fractional charge of $-1/8$.
In this configuration, the total dipole moment is zero, and consequently, the boundary contribution to the total unit cell energy vanishes.
However, the boundary term still yields a nonzero contribution to the electrostatic potential experienced by the central ion. 

\begin{table}[!htb]   %%%%%%%%%%%%%%%
\caption{Madelung constants ($\sqrt{3}\nu_{\rm pbc}/2$) for CsCl (${\mathbf r}=(0.5, 0.5, 0.5)$) computed using Eq.~\eqref{eq:pbc}, where $-\sqrt{3}\nu_{\rm b}/2=\sqrt{3}\pi/4 \simeq 1.36$ from Eq.~\eqref{eq:bc} and $\nu_{\rm corr}$ is from Eq.~\eqref{eq:corr}. 
With this method, the boundary contribution dominates the Madelung constant. The reference value is 1.76\,267\,477\,307\,098\cite{csclseq}.}

\label{tab:cscl}
\begin{tabular}{cclcc}\hline $ p$ & $\sqrt{3}\nu_{\rm pbc}/2$    &  $\,$ abs. error ($\epsilon$)  $\,$  & $\sqrt{3}\nu/2$ & $-\sqrt{3}\nu_{\rm corr}/2$  \\[0.4ex] \hline
                              1   &               1.7629780255   & \,\,3.0 $\times 10^{-4}$             &  0.439666       &  -3.7 $\times 10^{-2}$       \\[0.5ex]
                              5   &               1.7626721815   &    -2.6 $\times 10^{-6}$             &  0.405077       &  -2.8 $\times 10^{-3}$       \\[0.5ex]
                             20   &               1.7626747599   &    -1.3 $\times 10^{-8}$             &  0.402524       &  -2.0 $\times 10^{-4}$       \\[0.5ex]
                             60   &               1.7626747729   &    -1.7 $\times 10^{-10}$            &  0.402348       &  -2.3 $\times 10^{-5}$       \\\hline
\end{tabular}\end{table}
By isolating the bulk contribution from the finite-lattice sum, we obtain:
\begin{equation} \nu_{\rm pbc}({\mathbf r}) = \nu({\mathbf r},p|{\mathbf s}) - \nu_{\rm b}({\mathbf r}|{\mathbf s}) - \nu_{\rm corr}({\mathbf r},p|{\mathbf s}) .  \label{eq:pbc} \end{equation}
This formula establishes an explicitly corrected direct-summation scheme for evaluating the bulk potential.
While integral-transform techniques (most notably Ewald summation and its variants\cite{Ewald1921,Nijboer1957,DeLeeuw_Smith1980,Borwein1985}) are highly accurate, conceptually simpler direct-summation approaches\cite{Evjen1932,Harrison2006,Marathe1983,Sousa1993,Derenzo2000,Gelle2008,Tavernier2020,Tavernier2021} remain attractive due to their physical transparency.
Remarkably, by explicitly correcting for both boundary and finite-size effects, Tab.~\ref{tab:cscl} shows that this scheme achieves an accuracy of $10^{-4}$ using only a minimal $3\times3\times3$ supercell for the Madelung constant of CsCl.
Clearly, when the displacement is much smaller than the unit cell length, even the $p=0$ result --- $1/r$ with boundary and finite-size effects removed --- can approximate the bulk term well.
Within this corrected scheme, even if the leading-order finite-size term is neglected, subtracting the boundary term alone yields a rapidly converging scheme with an error scaling of ${\mathcal O}(p^{-2})$.

As anticipated by the pairwise formulation in Eq.~\eqref{eq:phi}, comparable performance extends to more complex cubic structures like NaCl (rocksalt), ZnS (zincblende), CaF$_2$ (fluorite), and CaTiO$_3$~\cite{Zhao_Hu2026}.
%\textcolor{blue}{
Furthermore, the explicit expressions of the correction terms in Eqs.~\eqref{eq:corr} and~\eqref{eq:b} reveal that uncorrected direct summation can still yield accurate electrostatic potentials, provided the unit cell is chosen such that the boundary term---or even the leading-order finite-size correction---vanishes.
Tab.~\ref{tab:unit} lists the ion coordinates for such unit cells in typical rocksalt and perovskite structures.
As illustrated in Tab.~\ref{tab:naca}, for the Na$_4$Cl$_4$ unit cell, the uncorrect direct sum gives the bulk electrostatic potential of Na$^+$ to an accuracy of ${\mathcal O}(p^{-4})$ since both the quadratic and quartic contributions vanish\cite{Zhao_Hu2025,Zhao_Hu2026}.
For the CaTiO$_3$ unit cell, the quadratic contributions also vanish, but the quartic contribution remains; consequently, the uncorrected direct sum achieves an accuracy of ${\mathcal O}(p^{-2})$. 
%}
\begin{table}[!htb]  %%%%%%%%%%%%%%% table unit cell coordinates %%%%%%%%%%%%%%%%
%\textcolor{blue}{
\caption{Reduced coordinates of ions in a cubic unit cell for typical rocksalt (Na$_4$Cl$_4$, left two columns) and perovskite (CaTiO$_3$, right two columns) structures, with the unit cell scaled to unit length: $ l = 1 $.}\label{tab:unit}
\begin{tabular}{ll|ll}\hline $4$Na$^+$            &       $4$Cl$^-$                &\quad  Ca$^{2+}$ & (0.5, 0.5, 0.5) \\[0.5ex]
                             (0, 0, 0)            &       (0.5, 0.5, 0.5)          &\quad  Ti$^{4+}$ &  (0, 0, 0)      \\[0.5ex]
                        (0.5, 0.5, 0)  \quad\quad &       (0.5, 0, 0)              &\quad  O$^{2-}$  & (0.5, 0, 0)     \\[0.5ex]
                             (0, 0.5, 0.5)        &       (0, 0.5, 0)              &\quad  O$^{2-}$  & (0, 0.5, 0)     \\[0.5ex]
                             (0.5, 0, 0.5)        &       (0, 0, 0.5)              &\quad  O$^{2-}$  & (0, 0, 0.5)     \\\hline
\end{tabular}
%}
\end{table}
\begin{table*}[!htb]  %%%%%%%%%%%%%%% table uncorrected direct summation %%%%%%%%%%%%%%%%
%\textcolor{blue}{
\caption{Electrostatic potentials computed via uncorrected direct summation for the unit cells listed in Table~\ref{tab:unit}, along with the scaling behavior of the absolute error ($\epsilon$) as a function of the crystal size $p$ ($K=2p+1$).}\label{tab:naca}
\begin{tabular}{cccccccccc}\hline $p$ & $\quad K\quad$ & $\quad\quad$  & \qquad Na$^+$\qquad  &  $\epsilon$           & $\epsilon K^4$ & $\qquad$ & \qquad Ca$^{2+}$\qquad &      $\epsilon$      & $\epsilon K^2$ \\[0.5ex]
                                  1   & 3              & \quad         &        -1.7470415645 &  5.2$\times10^{-4}$   &  0.042         & \qquad   &      -2.629494         &  6.4$\times10^{-2}$  &    0.58        \\[0.5ex]
                                  2   & 5              & \quad         &        -1.7475005023 &  6.4$\times10^{-5}$   &  0.040         & \qquad   &      -2.670713         &  2.3$\times10^{-2}$  &    0.57        \\[0.5ex]
                                  5   & 11             & \quad         &        -1.7475618563 &  2.7$\times10^{-6}$   &  0.040         & \qquad   &      -2.688842         &  4.8$\times10^{-3}$  &    0.58        \\[0.5ex]
                                  10  & 21             & \quad         &        -1.7475643885 &  2.1$\times10^{-7}$   &  0.040         & \qquad   &      -2.692296         &  1.3$\times10^{-3}$  &    0.58        \\[0.5ex]
                                  20  & 41             & \quad         &        -1.7475645804 &  1.4$\times10^{-8}$   &  0.040         & \qquad   &      -2.693261         &  3.4$\times10^{-4}$  &    0.58        \\[0.5ex]
                                  40  & 81             & \quad         &        -1.7475645937 &  9.3$\times10^{-10}$  &  0.040         & \qquad   &      -2.693517         &  8.8$\times10^{-5}$  &    0.58        \\\hline
\end{tabular} 
%}
\end{table*}

%\textcolor{blue}{
The current formulation, which focuses on a centrosymmetric crystal with its central unit cell, provides a basis for treating either non-centrosymmetric crystals or centrosymmetric crystals with off-center unit cells.
Such systems can be decomposed into a centrosymmetric crystal and a few one- or two-dimensional crystals, for which the boundary terms are no longer conditional\cite{Hu2014ib}.
Extending the present calculation from the classical Madelung problem to the off-center Madelung problem remains an interesting objective for future research.
%}

%%%%%%%%%%%%%%%%%%%%%%%%%%%%%%%%%%%%%%%%%%%%%%%%%%%%%%%%%%%%%%%%%%%%%%%%%%%%%%%%%%%%%%%%%%%%%%%%%%%%%%%%%%%%%%%%%%%%%%%%%%%%%%%%%%%%%%%%%%%%%%%%%%%%%%%%%%%%%%%%  ------------------- boundary term 
\begin{tcolorbox}[colback=yellow!30, colframe=black!75,colframe=white,boxrule=0pt]
{\Large  Within the framework of symmetry-preserving mean-field theory, the concise pairwise expression for $\nu_{\rm pbc}$ enables transparent development of approximate algorithms for condensed matter simulations.}
\end{tcolorbox}
The effective pairwise interaction $\nu_{\rm pbc}$ (previously denoted as $\nu_{\rm e3dtf}$\cite{Yi_Hu2017pairwise,Zhao_Hu2025}) characterizes periodic electrostatic systems through eight fundamental properties: symmetry and positivity, lattice periodicity, dominance over the bare Coulomb interaction, cancellation of
electric-field at the surface of a simulation box, constant average potential, constant potential for a uniform
charge distribution, bulk invariance, and scaling behavior. 
%\textcolor{blue}{
These properties are discussed in detail in Section 3 of Ref.\cite{Zhao_Hu2025} for both the present Coulomb lattice sum and lattice sums based on non-Coulombic interaction.
%}
Its compact pairwise formulation has been widely adopted to analytically predict structural properties and dielectric responses within the framework of symmetry-preserving mean-field (SPMF) theory\cite{Hu2014spmf,Yi_Hu2017mf}, with successful applications to both interfacial\cite{Pan_Hu2017,Pan_Hu2019} and bulk systems\cite{Pan_Hu2017,Hu2022}.
While a comprehensive review of SPMF theory and related mean-field theories\cite{Waals1873,Debye_Huckel1923,Widom1967,Weeks2002,Hu_Weeks2010,Gao_Weeks2020,Gao_Remsing2022,Remsing2023} lies beyond the scope of this Perspective, we highlight one particularly promising application: integrating the SPMF formalism with widely used mesh-based lattice-sum algorithms to systematically correct grid-induced artifacts and enhance computational accuracy.

\begin{figure}[!htb]\centerline{\includegraphics[width=6cm]{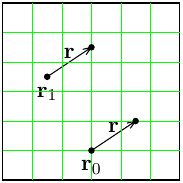} }          %---------------------  Figure -------------------------------------
\caption{Schematic of a unit cell (black box) with a $6\times 6$ mesh grid (green lines) and two charge pairs (dots) with identical separation ${\mathbf r}$ but different absolute positions relative to the mesh. 
The grid-dependent interaction $\nu_{\rm FG}$ violates translational invariance, yielding 
$\nu_{\rm FG}({\mathbf r}_1, {\mathbf r}+{\mathbf r}_1) \neq \nu_{\rm FG}({\mathbf r}_0, {\mathbf r}+{\mathbf r}_0)$.
\label{fig:mesh}}\end{figure}
In particle-mesh type algorithms (e.g.,\cite{Hockney_Eastwood1981,Darden1993,Essmann1995}), the Fourier-space contribution $\nu_{\rm F}({\mathbf r}_1 - {\mathbf r}_2)$ is approximated by a grid-interpolated function $\nu_{\rm FG}({\mathbf r}_1, {\mathbf r}_2)$. 
Unlike the exact Fourier term, this discretization breaks translational invariance, depending not only on the relative displacement ${\mathbf r}_1 - {\mathbf r}_2$ but also on the absolute positions of the particles relative to the mesh (see Fig.~\ref{fig:mesh}).
Consequently, the grid-based treatment introduces spurious self-forces\cite{Ballenegger2011} and can produce significant errors in short-range charge-charge interactions, particularly when coarse grid spacings are employed.

\begin{figure}[!htb]\centerline{\includegraphics[width=7cm]{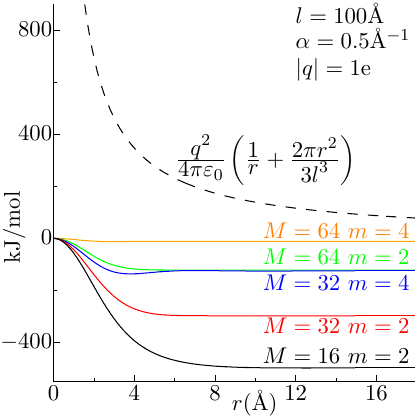} }          %---------------------  Figure -------------------------------------
%\textcolor{blue}{
\caption{
Averaged electrostatic interaction from the Ewald method (dashed line) in a cubic simulation box ($l=100$\AA), and its difference (solid lines) from a particle-mesh treatment with mesh size $M^3$, interpolation order $m$, and screening parameter $\alpha=0.5$\AA.
This interaction corresponds to the negative of the electrostatic energy between a pair of unit charges. The prefactor $1/(4\pi \varepsilon_0)$ is included to give energies in units of kJ/mol.}
\label{fig:uadd} 
%}
\end{figure}

Within the SPMF framework, these mesh-induced artifacts can be systematically corrected by introducing an isotropic short-range correction term $\nu_{\rm add}$ defined as
\begin{equation} \nu_{\rm add}(r) = \left< \nu_{\rm pbc}({\mathbf r}) \right>_{\rm sp}  - \left< \nu_{\rm FG}({\mathbf r}_1, {\mathbf r}+{\mathbf r}_1) + \nu_{\rm R}({\mathbf r}) \right>_{\rm sp}, \label{eq:nuadd} \end{equation}
where the symmetry-preserving average acting on an arbitrary function $f({\mathbf r}_1, {\mathbf r})$ is given by
\begin{equation}\left< f({\mathbf r}_1,{\mathbf r})\right>_{\rm sp}  = \int_V d{\mathbf r}_1 \int_0^{2\pi}d\phi\int_0^\pi \sin\theta d\theta\frac{ f({\mathbf r}_1,{\mathbf r})  }{4\pi V} . \end{equation}
Here, $\theta$ and $\phi$ denote the polar and azimuthal angles of ${\mathbf r}$, respectively, and $V$ is the volume of the unit cell (simulation box).
The average over the exact bulk term is analytically known:  $\left< \nu_{\rm pbc}({\mathbf r}) \right>_{\rm sp} = 1/r - 2\pi r^2/(3V)$\cite{Zhao_Hu2025}.
By adding $\nu_{\rm add}(r)$, which can be easily handled together with $\nu_{\rm R}({\mathbf r})$ in Eq.~\eqref{eq:er}, the overall interaction on average restores the short-range interaction prescribed by $\nu_{\rm pbc}$ without altering its long-range behavior.
%\textcolor{blue}{
Fig.~\ref{fig:uadd} displays typical $\nu_{\rm add}(r)$ for a model system of ions in a cubic simulation box ($l=100$\AA) with electrostatics handled by the smooth particle-mesh algorithm\cite{Essmann1995} using a mesh size of $M\times M\times M$, an interpolation order of $m$, and a screening parameter $\alpha= 0.5$\AA. 
Since $\nu_{\rm add}(r)$ represents the difference in electrostatic energy for a pair of unit charges (e.g., Na$^+$ and Cl$^-$), including $\nu_{\rm add}(r)$ ensures that the interaction on average is properly accounted for, even when smaller $M$ and $m$ are used.
Although $\nu_{\rm add}(r)$ produces short-ranged forces, its slowly varying behavior compared to the overall electrostatic forces justifies the mean-field approximation.
%}

While the correction scheme from SPMF is theoretically well-founded, its practical performance when integrated with established mesh-based algorithms --- from the widely used earlier PME method\cite{Essmann1995} to more recent improvements for bulk\cite{Shamshirgar_Tornberg2021,Liang2026} and interfacial simulations\cite{Ahrens-Iwers2021,Huang_Gao2024} --- remains an open question.
Notably, analogous SPMF-based corrections have recently been successfully applied to eliminate long-range artifacts in reciprocal space\cite{Hu2022,Gao_Hu2023,Gao2026},
suggesting that extending this formalism to general particle-mesh schemes is a promising avenue for improving both accuracy and computational efficiency in condensed matter simulations.

%%%%%%%%%%%%%%%%%%%%%%%%%%%%%%%%%%%%%%%%%%%%%%%%%%%%%%%%%%%%%%%%%%%%%%%%%%%%%%%%%%%%%%%%%%%%%%%%%%%%%%%%%%%%%%%%%%%%%%%%%%%%%%%%%%%%%%%%%%%%%%%%%%%%%%%%%%%%%%%%
\section*{Appendix: Lattice Summation for Arbitrarily Shaped Unit Cells}\label{sec:appdx}
\renewcommand{\theequation}{A\arabic{equation}}  % ← key line
\setcounter{equation}{0}                          % restart numbering

In this appendix, we provide the electrostatic energies of a periodic system with an arbitrary unit cell geometry.  
Consider a primary unit cell containing $N$ point charges (with charge $q_j$ at positions ${\mathbf r}_j$, $j=1,2,\cdots N$) and one continuous charge density $\rho({\mathbf r})$.
The lattice is defined by primitive vectors ${\mathbf a}_\mu$ ($\mu =1,2,3$), which form the edges of the unit cell with positive volume $V=\left( {\mathbf a}_1 \times {\mathbf a}_2\right)\cdot {\mathbf a}_3$.
The corresponding reciprocal lattice vectors ${\mathbf b}_\mu$ are defined by
%\[\frac{ {\mathbf b}_1 }{2\pi} = \frac{{\mathbf a}_2\times{\mathbf a}_3}{V}\quad \mbox{etc.} \]
\begin{equation} \frac{ {\mathbf b}_1 }{2\pi} = \frac{{\mathbf a}_2\times{\mathbf a}_3}{V};\quad \frac{ {\mathbf b}_2 }{2\pi} = \frac{{\mathbf a}_3\times{\mathbf a}_1}{V};\quad\frac{ {\mathbf b}_3 }{2\pi} = \frac{{\mathbf a}_1\times{\mathbf a}_2}{V}, \end{equation}
such that ${\mathbf a}_\mu \cdot {\mathbf b}_\beta = 2\pi\delta_{\mu\beta}$ (the Kronecker delta), for $\mu,\beta = 1,2,3$.

In the Ewald formulation\cite{Ewald1921}, now commonly referred to as the three-dimensional Ewald summation associated with the tinfoil (conducting) boundary condition (e3dtf)\cite{Felderhof1980,DeLeeuw_Smith1980,Hu2014ib,Yi_Hu2017pairwise,Zhao_Hu2025}, the pairwise interaction 
\begin{equation}\nu_{\rm pbc}({\mathbf r}) \equiv \nu_{\rm e3dtf}({\mathbf r})= \nu_{\rm R}({\mathbf r}) + \nu_{\rm F}({\mathbf r}), \label{eq:e3dtf0} \end{equation}
is split into a near-field contribution, expressed as a real-space series:
\begin{equation} \nu_{\rm R}({\mathbf r}) = \sum_{\mathbf n}  \frac{ {\rm erfc}\left(\alpha \lvert {\mathbf r}+ {\mathbf n} \rvert\right)}{ \lvert {\mathbf r} + {\mathbf n} \rvert} -  \sum_{ {\mathbf n}\neq {\mathbf 0} }  \frac{ {\rm erfc}(\alpha \lvert {\mathbf n}\rvert ) }{\lvert {\mathbf n} \rvert } +
\frac{2\alpha}{\sqrt{\pi}}, \label{eq:er}  \end{equation}
and a far-field contribution, expressed as a Fourier-space series:
\begin{equation} \nu_{\rm F}({\mathbf r}) = \frac{4\pi}{V}\sum_{{\mathbf k}\neq {\mathbf 0} } \frac{e^{-k^2/(4\alpha^2)}}{k^2}\left(e^{i{\mathbf k}\cdot{\mathbf r}} - 1 \right), \label{eq:e3dtf2}  \end{equation}
where 
${\mathbf n} = n_1 {\mathbf a}_1 + n_2 {\mathbf a}_2 + n_3 {\mathbf a}_3$ and ${\mathbf k} = k_1 {\mathbf b}_1 + k_2 {\mathbf b}_2 + k_3 {\mathbf b}_3$ with $n_1$, $n_2$, $n_3$, $k_1$, $k_2$, $k_3$ all being integers. 
Here, ${\rm erfc}(x) = 1- {\rm erf}(x)$ is the complementary error function.
The summations in Eqs.~\eqref{eq:er} and~\eqref{eq:e3dtf2} extend over all integers triplets from $-\infty$ to $\infty$, with ${\mathbf n}\neq {\mathbf 0}$ indicating that the $n_1=n_2=n_3=0$ term is excluded.
These expressions reduce to those of earlier works\cite{Hu2014ib,Yi_Hu2017pairwise,Zhao_Hu2025} when the lattice vectors are orthogonal.
%\textcolor{blue}{
Of course, the Ewald splitting using the error function is not the only choice\cite{Hockney_Eastwood1981}. Very recently, an optimal splitting has been developed to minimize the error in the Fourier-space summation\cite{Liang2026}.
%}

By choosing the parameter $\alpha > 0$ appropriately, the computation of $\nu_{\rm pbc}({\mathbf r})$ via Eqs.~\eqref{eq:er} and~\eqref{eq:e3dtf2} becomes efficient for any ${\mathbf r}$.
%\textcolor{blue}{
The symmetry-preserving averages acting on the real- and Fourier-space series yield their spherical averages:
\begin{equation} 
\left< \nu_{\rm R}({\mathbf r}) \right>_{\rm sp} = \frac{ {\rm erfc}(\alpha r)}{r} + \frac{2\alpha}{\sqrt{\pi}}, \label{eq:aer}
\end{equation}
and
\begin{equation}
\left< \nu_{\rm F}({\mathbf r}) \right>_{\rm sp} = \frac{ {\rm erf}(\alpha r)}{r} + \frac{2\pi}{3V} r^2 - \frac{2\alpha}{\sqrt{\pi}}, \label{eq:aef}
\end{equation}
for ${\mathbf r}$ within the simulation box.
In particle-mesh type treatments (e.g.,\cite{Essmann1995}), $\nu_{\rm F}({\mathbf r})$ is approximated by a grid-based function $\nu_{\rm FG}({\mathbf r}_1, {\mathbf r}_1 + {\mathbf r})$ that is no longer translationally invariant.
Nevertheless, the exact condition $\nu_{\rm F}({\mathbf 0})=0$ can be guaranteed by enforcing that the sum of the coefficients in the discrete Fourier transform vanishes.
The approximation away from ${\mathbf r}={\mathbf 0}$ after averaging (see Eq.~\eqref{eq:nuadd}),  
\begin{equation} \nu_{\rm add}(r) = \left<\nu_{\rm F}({\mathbf r}) \right>_{\rm sp} - \left<\nu_{\rm FG}({\mathbf r}_1, {\mathbf r}_1 + {\mathbf r}) \right>_{\rm sp}, \end{equation}
is shown in Fig.~\ref{fig:uadd}.
%}

Alternatively, the pairwise interaction can be expressed entirely as a Fourier-space series\cite{Yi_Hu2017pairwise,Zhao_Hu2025}:
\begin{equation} \nu_{\rm pbc}({\mathbf r}) = \nu_{\rm R}({\mathbf r}) +  \nu_{\rm F}({\mathbf r}) = \tau + \frac{4\pi}{V}\sum_{{\mathbf k}\neq {\mathbf 0} } \frac{e^{-i{\mathbf k}\cdot {\mathbf r}}}{k^2},  \end{equation}
where $\tau$ is a geometry-dependent constant
\begin{equation} \tau = \frac{\pi}{\alpha^2V} + \frac{2\alpha}{\sqrt\pi} -  \sum_{ {\mathbf n}\neq {\mathbf 0} }  \frac{ {\rm erfc}(\alpha \lvert {\mathbf n}\rvert ) }{\lvert {\mathbf n} \rvert } -\frac{4\pi}{V}\sum_{{\mathbf k}\neq {\mathbf 0} }\frac{e^{-k^2/(4\alpha^2)}}{k^2}. \label{eq:tau} \end{equation}
Physically, $\tau$ corresponds to the bulk potential averaged over the unit cell, generated by a unit point charge, or equivalently, the potential at an arbitrary point due to a uniform charge density $1/V$ under the PBC\cite{Zhao_Hu2025}.
This expression generalizes Eq.(5) of Ref.\cite{Yi_Hu2017pairwise} to arbitrary triclinic lattices. In the special case of a cubic lattice ($V=l^3$), $\tau$ simplifies to $2.83729748/l$ [cf. Eq.(28) in Ref.\cite{Zhao_Hu2025}]. 
Notably, although $\alpha$ appears explicitly in Eq.~\eqref{eq:tau}, $\tau$ is strictly independent of the splitting parameter and is determined solely by the lattice geometry.

\begin{tcolorbox}[colback=yellow!30, colframe=black!75,colframe=white,boxrule=0pt]
{\Large The pairwise formulation gives the transparent derivations of the Ewald sum energies and pressure for mixed discrete-continous charges of non-neutral systems.}
\end{tcolorbox}
The introduction of the pairwise form enables a concise expression of the total electrostatic energy for the system with mixed discrete and continuous charges:
\begin{equation} {\mathcal U} = {\mathcal U}_{\rm pp} + {\mathcal U}_{\rm pc} +  {\mathcal U}_{\rm cc}, \end{equation}
where ${\mathcal U}_{\rm pp}$, ${\mathcal U}_{\rm pp}$ and ${\mathcal U}_{\rm pp}$ represent the particle-particle (pp), particle-continuum (pc) and continuum-continuum (cc) interaction energies, respectively,
\begin{equation} {\mathcal U}_{\rm pp} = \sum_{i<j}q_iq_j \nu_{\rm pbc}({\mathbf r}_i - {\mathbf r}_j), \label{eq:pp} \end{equation}
\begin{equation} {\mathcal U}_{\rm pc} = \sum_{j=1}^N q_j  \int d{\mathbf r}\, \rho({\mathbf r}) \nu_{\rm pbc}({\mathbf r}-{\mathbf r}_j)   , \label{eq:pc} \end{equation}
and
\begin{equation} {\mathcal U}_{\rm cc} = \frac{1}{2} \int d{\mathbf r} \int d{\mathbf r}^\prime\, \rho({\mathbf r}) \rho({\mathbf r}^\prime)  \nu_{\rm pbc}({\mathbf r}-{\mathbf r}^\prime) . \label{eq:cc} \end{equation}
The pairwise form of ${\mathcal U}_{\rm pp}$ differs from the conventional Ewald sum (see e.g. Eq. (6.2) of Ref.\cite{Allen_Tildesley2017} and Eq. (11.2.24) of Ref.\cite{Frenkel_Smit2023})
%\begin{multline} {\mathcal U}_0= \frac{1}{2}\sum_{i,j=1}^N q_iq_j\sideset{}{'}\sum_{\mathbf n} \frac{{\rm erfc}(\alpha|{\mathbf r}_{ij}+{\mathbf n}|)} {|{\mathbf r}_{ij}+{\mathbf n}|} - \frac{\alpha}{\sqrt\pi} \sum_{j=1}^N q_j^2 \\
%+\frac{2\pi}{V}\sum_{i,j=1}^N q_iq_j\sum_{{\mathbf k}\neq{\mathbf 0}} \frac{e^{-k^2/(4\alpha^2)}}{k^2}  e^{i{\mathbf k}\cdot{\mathbf r}_{ij}}    \end{multline}
\begin{multline} {\mathcal U}_0= \frac{1}{2}\sum_{i,j=1}^N q_iq_j\sideset{}{'}\sum_{\mathbf n} \frac{{\rm erfc}(\alpha|{\mathbf r}_{ij}+{\mathbf n}|)}
                {|{\mathbf r}_{ij}+{\mathbf n}|} - \frac{\alpha}{\sqrt\pi} \sum_{j=1}^N q_j^2 \\
+\frac{2\pi}{V}\sum_{{\mathbf k}\neq{\mathbf 0}} \frac{e^{-k^2/(4\alpha^2)}}{k^2}\left| \sum_{j=1}^N  e^{i{\mathbf k}\cdot{\mathbf r}_j} \right|^2, \label{eq:ew}
   \end{multline}
by a term that depends quadratically on the total discrete charge $Q=\sum_{j=1}^N q_j$:
\begin{equation} {\mathcal U}_{\rm pp}\equiv \sum_{i<j}q_iq_j \nu_{\rm pbc}({\mathbf r}_{ij}) = {\mathcal U}_0 +\frac{Q^2}{2} \left[ \tau - \frac{\pi}{\alpha^2 V} \right]. \label{eq:upp2} \end{equation}
In Eq.~\eqref{eq:ew}, the prime denotes that terms with ${\mathbf n}={\mathbf 0}$ and $i=j$ are omitted. To derive Eq.~\eqref{eq:upp2}, we have frequently applied the identity
\begin{equation} \sum_{i<j} q_i q_j f({\mathbf r}_{ij}) = \frac{1}{2}\sum_{i,j=1}^N  q_i q_j f({\mathbf r}_{ij}) - \frac{1}{2} \sum_{j=1}^N q_j^2 f({\mathbf 0}), \end{equation}
where $f({\mathbf r})$ is an arbitrary even function of ${\mathbf r}$.
While the pairwise expression ${\mathcal U}_{\rm pp}$ is rigorously independent of the splitting parameter $\alpha$, the conventional expression ${\mathcal U}_0$ exhibits a spurious $\alpha$-dependence whenever the system is non-neutral ($Q\neq 0$)\cite{Yi_Hu2017pairwise}. 

For the specific case of a uniform neutralizing background\cite{Onegin2024,Li2025,Demyanov2025}, $\rho({\mathbf r}) = -Q/V$, the particle-continuum and continuum-continuum contributions reduce to particularly simple forms: ${\mathcal U}_{\rm pc} = - 2 {\mathcal U}_{\rm cc} = - \tau Q^2$. This relation is fully consistent with the physical interpretation of $\tau$. Consequently, the total electrostatic energy becomes\cite{Zhao_Hu2025}
\begin{equation} {\mathcal U}_{\rm homo} \equiv {\mathcal U}_{\rm pp} - \frac{\tau Q^2}{2} = {\mathcal U}_0 - \frac{\pi Q^2}{ 2 \alpha^2 V}.  \end{equation}
%%%%%%%%%%%%%%%%%%%%%%%%%%%%%%%%%%%%%%%%%%%%%%%%%%%%%%%%%%%%%%%%%%%%%%%%%%%%%%%%%%%%%%%%%%%%%%%%%%%%%%%%%%%%%%%%%%%%%%%%%%%%%%%%%%%%%%%%%%%%%%%%%%%%%%%%%%%%%%%% 
\section*{Acknowledgement}
This work was supported by NSFC (Grant Nos. 22273047 and 21873037) and Shandong Provincial Special Zone for Fundamental Research (Chemistry) (Grant No. TQ022025003).
On the occasion of the 80th anniversary of Jilin University, Z. H. greatly acknowledges the longstanding support from the university and the chemistry department for our initial work on lattice summations\cite{Pan_Hu2014,Hu2014ib,Pan_Hu2015,Yi_Hu2017pairwise,Pan_Hu2019} and the symmetry-preserving mean-field theory\cite{Hu2014spmf,Yi_Hu2015,Yi_Hu2017mf,Pan_Hu2017}.

%\bibliography{refelec}

%merlin.mbs apsrev4-1.bst 2010-07-25 4.21a (PWD, AO, DPC) hacked
%Control: key (0)
%Control: author (0) dotless jnrlst
%Control: editor formatted (1) identically to author
%Control: production of article title (0) allowed
%Control: page (1) range
%Control: year (0) verbatim
%Control: production of eprint (0) enabled
%
\end{document}